\catcode`\@=11

\font\tenmsa=msam10
\font\sevenmsa=msam7
\font\fivemsa=msam5
\font\tenmsb=msbm10
\font\sevenmsb=msbm7
\font\fivemsb=msbm5
\newfam\msafam
\newfam\msbfam
\textfont\msafam=\tenmsa  \scriptfont\msafam=\sevenmsa
  \scriptscriptfont\msafam=\fivemsa
\textfont\msbfam=\tenmsb  \scriptfont\msbfam=\sevenmsb
  \scriptscriptfont\msbfam=\fivemsb

\def\hexnumber@#1{\ifnum#1<10 \number#1\else
 \ifnum#1=10 A\else\ifnum#1=11 B\else\ifnum#1=12 C\else
 \ifnum#1=13 D\else\ifnum#1=14 E\else\ifnum#1=15 F\fi\fi\fi\fi\fi\fi\fi}

\def\msa@{\hexnumber@\msafam}
\def\msb@{\hexnumber@\msbfam}
\mathchardef\boxdot="2\msa@00
\mathchardef\boxplus="2\msa@01
\mathchardef\boxtimes="2\msa@02
\mathchardef\square="0\msa@03
\mathchardef\blacksquare="0\msa@04
\mathchardef\centerdot="2\msa@05
\mathchardef\lozenge="0\msa@06
\mathchardef\blacklozenge="0\msa@07
\mathchardef\circlearrowright="3\msa@08
\mathchardef\circlearrowleft="3\msa@09
\mathchardef\rightleftharpoons="3\msa@0A
\mathchardef\leftrightharpoons="3\msa@0B
\mathchardef\boxminus="2\msa@0C
\mathchardef\Vdash="3\msa@0D
\mathchardef\Vvdash="3\msa@0E
\mathchardef\vDash="3\msa@0F
\mathchardef\twoheadrightarrow="3\msa@10
\mathchardef\twoheadleftarrow="3\msa@11
\mathchardef\leftleftarrows="3\msa@12
\mathchardef\rightrightarrows="3\msa@13
\mathchardef\upuparrows="3\msa@14
\mathchardef\downdownarrows="3\msa@15
\mathchardef\upharpoonright="3\msa@16

\mathchardef\downharpoonright="3\msa@17
\mathchardef\upharpoonleft="3\msa@18
\mathchardef\downharpoonleft="3\msa@19
\mathchardef\rightarrowtail="3\msa@1A
\mathchardef\leftarrowtail="3\msa@1B
\mathchardef\leftrightarrows="3\msa@1C
\mathchardef\rightleftarrows="3\msa@1D
\mathchardef\Lsh="3\msa@1E
\mathchardef\Rsh="3\msa@1F
\mathchardef\rightsquigarrow="3\msa@20
\mathchardef\leftrightsquigarrow="3\msa@21
\mathchardef\looparrowleft="3\msa@22
\mathchardef\looparrowright="3\msa@23
\mathchardef\circeq="3\msa@24
\mathchardef\succsim="3\msa@25
\mathchardef\gtrsim="3\msa@26
\mathchardef\gtrapprox="3\msa@27
\mathchardef\multimap="3\msa@28
\mathchardef\therefore="3\msa@29
\mathchardef\because="3\msa@2A
\mathchardef\doteqdot="3\msa@2B

\mathchardef\triangleq="3\msa@2C
\mathchardef\precsim="3\msa@2D
\mathchardef\lesssim="3\msa@2E
\mathchardef\lessapprox="3\msa@2F
\mathchardef\eqslantless="3\msa@30
\mathchardef\eqslantgtr="3\msa@31
\mathchardef\curlyeqprec="3\msa@32
\mathchardef\curlyeqsucc="3\msa@33
\mathchardef\preccurlyeq="3\msa@34
\mathchardef\leqq="3\msa@35
\mathchardef\leqslant="3\msa@36
\mathchardef\lessgtr="3\msa@37
\mathchardef\backprime="0\msa@38
\mathchardef\risingdotseq="3\msa@3A
\mathchardef\fallingdotseq="3\msa@3B
\mathchardef\succcurlyeq="3\msa@3C
\mathchardef\geqq="3\msa@3D
\mathchardef\geqslant="3\msa@3E
\mathchardef\gtrless="3\msa@3F
\mathchardef\sqsubset="3\msa@40
\mathchardef\sqsupset="3\msa@41
\mathchardef\trianglerighteq="3\msa@44
\mathchardef\trianglelefteq="3\msa@45
\mathchardef\bigstar="0\msa@46
\mathchardef\between="3\msa@47
\mathchardef\blacktriangledown="0\msa@48
\mathchardef\blacktriangleright="3\msa@49
\mathchardef\blacktriangleleft="3\msa@4A
\mathchardef\blacktriangle="0\msa@4E
\mathchardef\triangledown="0\msa@4F
\mathchardef\eqcirc="3\msa@50
\mathchardef\lesseqgtr="3\msa@51
\mathchardef\gtreqless="3\msa@52
\mathchardef\lesseqqgtr="3\msa@53
\mathchardef\gtreqqless="3\msa@54
\mathchardef\Rrightarrow="3\msa@56
\mathchardef\Lleftarrow="3\msa@57
\mathchardef\veebar="2\msa@59
\mathchardef\barwedge="2\msa@5A
\mathchardef\doublebarwedge="2\msa@5B
\mathchardef\angle="0\msa@5C
\mathchardef\measuredangle="0\msa@5D
\mathchardef\sphericalangle="0\msa@5E
\mathchardef\varpropto="3\msa@5F
\mathchardef\smallsmile="3\msa@60
\mathchardef\smallfrown="3\msa@61
\mathchardef\Subset="3\msa@62
\mathchardef\Supset="3\msa@63
\mathchardef\Cup="2\msa@64

\mathchardef\Cap="2\msa@65

\mathchardef\curlywedge="2\msa@66
\mathchardef\curlyvee="2\msa@67
\mathchardef\leftthreetimes="2\msa@68
\mathchardef\rightthreetimes="2\msa@69
\mathchardef\subseteqq="3\msa@6A
\mathchardef\supseteqq="3\msa@6B
\mathchardef\bumpeq="3\msa@6C
\mathchardef\Bumpeq="3\msa@6D
\mathchardef\lll="3\msa@6E

\mathchardef\ggg="3\msa@6F

\mathchardef\circledS="0\msa@73
\mathchardef\pitchfork="3\msa@74
\mathchardef\dotplus="2\msa@75
\mathchardef\backsim="3\msa@76
\mathchardef\backsimeq="3\msa@77
\mathchardef\complement="0\msa@7B
\mathchardef\intercal="2\msa@7C
\mathchardef\circledcirc="2\msa@7D
\mathchardef\circledast="2\msa@7E
\mathchardef\circleddash="2\msa@7F
\def\ulcorner{\delimiter"4\msa@70\msa@70 }
\def\urcorner{\delimiter"5\msa@71\msa@71 }
\def\llcorner{\delimiter"4\msa@78\msa@78 }
\def\lrcorner{\delimiter"5\msa@79\msa@79 }
\def\yen{\mathhexbox\msa@55 }
\def\checkmark{\mathhexbox\msa@58 }
\def\circledR{\mathhexbox\msa@72 }
\def\maltese{\mathhexbox\msa@7A }
\mathchardef\lvertneqq="3\msb@00
\mathchardef\gvertneqq="3\msb@01
\mathchardef\nleq="3\msb@02
\mathchardef\ngeq="3\msb@03
\mathchardef\nless="3\msb@04
\mathchardef\ngtr="3\msb@05
\mathchardef\nprec="3\msb@06
\mathchardef\nsucc="3\msb@07
\mathchardef\lneqq="3\msb@08
\mathchardef\gneqq="3\msb@09
\mathchardef\nleqslant="3\msb@0A
\mathchardef\ngeqslant="3\msb@0B
\mathchardef\lneq="3\msb@0C
\mathchardef\gneq="3\msb@0D
\mathchardef\npreceq="3\msb@0E
\mathchardef\nsucceq="3\msb@0F
\mathchardef\precnsim="3\msb@10
\mathchardef\succnsim="3\msb@11
\mathchardef\lnsim="3\msb@12
\mathchardef\gnsim="3\msb@13
\mathchardef\nleqq="3\msb@14
\mathchardef\ngeqq="3\msb@15
\mathchardef\precneqq="3\msb@16
\mathchardef\succneqq="3\msb@17
\mathchardef\precnapprox="3\msb@18
\mathchardef\succnapprox="3\msb@19
\mathchardef\lnapprox="3\msb@1A
\mathchardef\gnapprox="3\msb@1B
\mathchardef\nsim="3\msb@1C
\mathchardef\napprox="3\msb@1D
\mathchardef\nsubseteqq="3\msb@22
\mathchardef\nsupseteqq="3\msb@23
\mathchardef\subsetneqq="3\msb@24
\mathchardef\supsetneqq="3\msb@25
\mathchardef\subsetneq="3\msb@28
\mathchardef\supsetneq="3\msb@29
\mathchardef\nsubseteq="3\msb@2A
\mathchardef\nsupseteq="3\msb@2B
\mathchardef\nparallel="3\msb@2C
\mathchardef\nmid="3\msb@2D
\mathchardef\nshortmid="3\msb@2E
\mathchardef\nshortparallel="3\msb@2F
\mathchardef\nvdash="3\msb@30
\mathchardef\nVdash="3\msb@31
\mathchardef\nvDash="3\msb@32
\mathchardef\nVDash="3\msb@33
\mathchardef\ntrianglerighteq="3\msb@34
\mathchardef\ntrianglelefteq="3\msb@35
\mathchardef\ntriangleleft="3\msb@36
\mathchardef\ntriangleright="3\msb@37
\mathchardef\nleftarrow="3\msb@38
\mathchardef\nrightarrow="3\msb@39
\mathchardef\nLeftarrow="3\msb@3A
\mathchardef\nRightarrow="3\msb@3B
\mathchardef\nLeftrightarrow="3\msb@3C
\mathchardef\nleftrightarrow="3\msb@3D
\mathchardef\divideontimes="2\msb@3E
\mathchardef\varnothing="0\msb@3F
\mathchardef\nexists="0\msb@40
\mathchardef\mho="0\msb@66
\mathchardef\thorn="0\msb@67
\mathchardef\beth="0\msb@69
\mathchardef\gimel="0\msb@6A
\mathchardef\daleth="0\msb@6B
\mathchardef\lessdot="3\msb@6C
\mathchardef\gtrdot="3\msb@6D
\mathchardef\ltimes="2\msb@6E
\mathchardef\rtimes="2\msb@6F
\mathchardef\shortmid="3\msb@70
\mathchardef\shortparallel="3\msb@71
\mathchardef\smallsetminus="2\msb@72
\mathchardef\thicksim="3\msb@73
\mathchardef\thickapprox="3\msb@74
\mathchardef\approxeq="3\msb@75
\mathchardef\succapprox="3\msb@76
\mathchardef\precapprox="3\msb@77
\mathchardef\curvearrowleft="3\msb@78
\mathchardef\curvearrowright="3\msb@79
\mathchardef\digamma="0\msb@7A
\mathchardef\varkappa="0\msb@7B
\mathchardef\hslash="0\msb@7D
\mathchardef\hbar="0\msb@7E
\mathchardef\backepsilon="3\msb@7F
\def\Bbb{\ifmmode\let\next\Bbb@\else
 \def\next{\errmessage{Use \string\Bbb\space only in math mode}}\fi\next}
\def\Bbb@#1{{\Bbb@@{#1}}}
\def\Bbb@@#1{\fam\msbfam#1}

\catcode`\@=\active

\def\CR{\hbox{{$\cal R$}}}

\def\R{{\Bbb R}}
\def\C{{\Bbb C}}

\def\vect{{\bf t}}\def\vecs{{\bf s}}\def\vecv{{\bf v}}
\def\vecu{{\bf u}}\def\vecx{{\bf x}}\def\vecp{{\bf p}}

\def\<{\langle}
\def\>{\rangle}
\def\del{{\partial}}

\def\eps{{\epsilon}}

\def\rcross{{\triangleright\!\!\!<}}

\def\dcross{{\bowtie}}

\def\tens{\mathop{\otimes}}
\def\ra{{\triangleleft}}
\def\isom{{\cong}}

\def\und#1{{\underline {#1}}}

\def\o{{}_{\scriptscriptstyle(1)}}
\def\t{{}_{\scriptscriptstyle(2)}}

\def\note#1{}

\def\equad{\kern -1.7em}

\def\nquad{{\!\!\!\!\!\!}}
\def\nqquad{\nquad\nquad}\def\nqqquad{\nqquad\nquad}
\def\eqn#1#2{\begin{equation}#2\label{#1}\end{equation}}
\def\cmath#1{\[\begin{array}{c} #1 \end{array}\]}

\def\ceqn#1#2{\begin{equation}
\label{#1}\begin{array}{c}#2\end{array}\end{equation}}

\def\coadd{{\Delta_+}}
\def\comult{{\Delta_\bullet}}
\def\Rrel{{\bf R}}
\def\Rmult{{\bf R_\bullet}}
\def\Radd{{\bf R_+}}
\def\dila{{\varsigma}} 
\def\mink{{\R_q^{1,3}}}

\documentstyle[11pt]{article}
\begin{document}\baselineskip 19pt

{\ }\hskip 4.7in DAMTP/93-68 
\vspace{.2in}

\begin{center} {\LARGE BRAIDED MATRIX STRUCTURE OF $q$-MINKOWSKI SPACE AND
$q$-POINCARE GROUP}
\\ \baselineskip 13pt{\ }
{\ }\\ S. Majid\footnote{Royal Society University Research Fellow and Fellow of
Pembroke College, Cambridge} \& U. Meyer\\ {\ }\\
Department of Applied Mathematics \& Theoretical Physics\\ University of
Cambridge, Cambridge CB3 9EW, U.K.
\end{center}

\begin{center}
December, 1993\end{center}
\vspace{10pt}
\begin{quote}\baselineskip 13pt
\noindent{\bf Abstract} We clarify the relation between the approach to
$q$-Minkowski space of Carow-Watamura {\em et al.} with an approach based on
the idea of $2\times 2$ braided Hermitean matrices. The latter are objects like
super-matrices but with Bose-Fermi statistics replaced by braid statistics. We
also obtain new R-matrix formulae for the $q$-Poincar\'e group in this
framework.
\end{quote}
\baselineskip 19.7pt

\section{Introduction}

This paper is a comment on the approach to $q$-Minkowski space developed in
\cite{CWSSW:ten}\cite{CWSSW:lor}\cite{OSWZ:def} based on the idea of the spinor
decomposition of Minkowski space vectors. From this point of view there was
found a suitable algebra of (non-commuting) co-ordinate functions as well as a
$q$-Lorentz quantum group associated to them.

Meanwhile, since 1990 a systematic approach to $q$-deformations has been
developed by the first author (S.M.) based on his idea of a {\em braided
matrix}. This is like a super-matrix but with the usual $\pm 1$ statistics
between co-ordinates replaced by braid statistics, typically controlled by an
R-matrix. A fairly complete theory of braided geometry is available by now,
modelled on the ideas of supersymmetry, and we refer to \cite{Ma:introp} for a
review of 30--40 papers on this topic.
The main idea in this approach is that $q$ should be viewed as a deformation of
the notion of $\tens$, i.e. it expresses how two independent copies of a system
fail to commute. It is obvious that the $2\times 2$ braided matrices introduced
in \cite{Ma:exa} could be taken as a definition of $q$-Minkowski space from
this point of view\cite{Ma:mec}\cite{Mey:new}. There are {\em three} $16\times
16$ R-matrices in this approach and we recall them in Section~2.

Here we map the $q$-Minkowski space and $q$-Lorentz group of
\cite{CWSSW:ten}--\cite{OSWZ:def} into this braided setting. Part of this
identification is clear because the algebra for $q$-Minkowski space there is
easily seen to be similar to that of $2\times 2$ braided Hermitean matrices.
But the underlying $16\times 16$ R-matrices and $q$-Lorentz groups appear at
first sight quite different in the two approaches (and were obtained quite
differently), requiring a lot of care to sort out. We do this in Section~3.

In Section~4 we use the braided theory to obtain a new braided-matrix
$q$-Poincar\'e group as a construction that goes beyond
\cite{CWSSW:ten}--\cite{OSWZ:def}. Indeed, the 30--40 papers on braided
geometry mentioned above can all be applied at once to $q$-Minkowski space in
this form. For this reason, we believe the detailed identification provided in
the present paper to be an important one.

\section{Braided-matrix approach}

A {\em braided matrix} as introduced by the first author in \cite{Ma:exa} means
an algebra with a matrix of generators $\vecu=\pmatrix{a&b\cr c&d}$ say, which
can be multiplied in the sense that
\[ \pmatrix{a''&b''\cr c''&d''}=\pmatrix{a&b\cr c&d}\pmatrix{a'&b'\cr c'&d'}\]
obeys the same relations as $a,b,c,d$ and another identical copy $a',b',c',d'$
provided (which is the new idea) the two copies are treated with suitable braid
statistics between them.

The $2\times 2$ braided matrices we need are the algebra $BM_q(2)$\cite{Ma:exa}
\cmath{ba=q^2ab,\quad ca=q^{-2}ac,\quad da=ad,\qquad
bc=cb+(1-q^{-2})a(d-a)\\
db=bd+(1-q^{-2})ab,\quad cd=dc+(1-q^{-2})ca}
and the {\em multiplicative braid statistics} between them was found also in
\cite{Ma:exa}
\eqn{bm2stat}{\begin{array}{rlll}
\nqquad&a'  a=a a'+(1-q^2)b c,' &a'  b=b  a,'& \nqqquad\nqquad a'  c=c
a'+(1-q^2)(d-a)  c'\\
\nqquad&a'  d=d  a'+(1-q^{-2})b  c,' & b'  a=a  b'+(1-q^2)b  (d'- a'), &b'
b=q^2b  b',\quad{\rm etc.}
\end{array}}
In all there are $16$ such relations. In particular, we showed in \cite{Ma:exa}
that
\[ t=q^{-1}a+qd,\quad \und{\rm det}=ad-q^2cb\]
are central and {\em bosonic} in the sense
\eqn{bosonic}{ x't=tx',\quad x'\und{\rm det}=\und{\rm det}x'\quad\forall x\in
BM_q(2).}

A {\em braided (co)vector space} as introduced by the first author in
\cite{Ma:poi} means an algebra with a (co)vector of generators $\vecx$ say,
which can be added in the sense that $\vecx''=\vecx+\vecx'$
obeys the same relations as $\vecx$ and another identical copy $\vecx'$
provided again that the two copies are treated with suitable braid statistics
between them. This applies for example to the quantum plane
$\vecx=(x,y)$\cite{Ma:poi}. The $2\times 2$ braided matrices above have this
additive structure too as found by the second author (U.M.). Thus,
\[  \pmatrix{a''&b''\cr c''&d''}=\pmatrix{a&b\cr c&d}+\pmatrix{a'&b'\cr
c'&d'}\]
obeys the same relations provided now that the two copies have the {\em
additive braid statistics}\cite{Mey:new}
\eqn{baddstat}{\begin{array}{lll}
a'  a=a a',&\nqquad\nqquad \nquad a'  b=q^{-2}b  a',&\nqquad\nquad \nqquad a'
c=c
a'+(1-q^{-2})ac',\\
a'  d=q^{-2}d  a'+(1-q^{-2})b  c' +(1-q^{-2})^2aa',& b'  a=a
b'+(1-q^{-2})ba',&b'
b=b  b',\quad {\rm etc.}
\end{array}}
Again there are 16 of these and (\ref{bosonic}) holds for our central element
$\und{\rm det}$, although not for $t$.

There is also a general R-matrix construction for $n\times n$ braided matrices
$B(R)$ and their properties\cite{Ma:exa}\cite{Mey:new}, with the above as an
example when one takes the $sl_2$ R-matrix
\eqn{Rjones}{R=\pmatrix{q&0&0&0\cr 0&1&q-q^{-1}&0\cr 0&0&1&0\cr
0&0&0&q}}
One can give the formulae directly in terms of such an $R$, or equivalently in
terms of the following three multi-index R-matrices
\eqn{Rrel}{\Rrel ^I{}_J{}^K{}_L=R^{-1}{}^{d}{}_{k_0}{}^{j_0}{}_{a}
R^{k_1}{}_{b}{}^{a}{}_{i_0}R^{i_1}{}_c{}^b{}_{l_1} {\widetilde
R}^c{}_{j_1}{}^{l_0}{}_d}
\eqn{Rmult}{\Rmult ^I{}_J{}^K{}_L=R^{j_0}{}_a{}^d{}_{k_0}
R^{-1}{}^a{}_{i_0}{}^{k_1}{}_b
R^{i_1}{}_c{}^b{}_{l_1} {\widetilde R}^c{}_{j_1}{}^{l_0}{}_d}
\eqn{Radd}{\Radd ^I{}_J{}^K{}_L=R^{j_0}{}_a{}^d{}_{k_0}
R^{k_1}{}_b{}^a{}_{i_0}
R^{i_1}{}_c{}^b{}_{l_1} {\widetilde R}^c{}_{j_1}{}^{l_0}{}_d}
where we used the multi-index notation $I=(i_0,i_1)$ etc, and where
$\widetilde{R}$ is given by transposition in the second two indices, inversion
and transposition again. The first two multi-index R-matrices were introduced
in \cite{Ma:exa} where they appear as $\Psi'$ and $\Psi$, while the third was
introduced in \cite{Mey:new}.

Then the braided matrices $B(R)$ as an algebra have the relations\cite{Ma:exa}
\eqn{B(R)}{
R_{21}\vecu_1R\vecu_2= \vecu_2 R_{21} \vecu_1 R,\quad {\rm i.e.}\quad
 u_Ju_L= u_Ku_I \Rrel ^I{}_J{}^K{}_L}
where the second form is with $u_I=u^{i_0}{}_{i_1}$ etc. This second vector
form is a general feature of these equations and it is for this reason that we
said in \cite{Ma:exa} that $B(R)$ is {\em braided-commutative} with respect to
$\Rrel$. We also studied the more familiar matrix form in
\cite{Ma:lin}\cite{Ma:skl} and explained its connection with the FRT
description of standard quantum groups with $\vecu=l^+Sl^-$. From this point of
view, (\ref{B(R)}) are the `quantum-Lie algebra'
relations\cite{ResSem:cen}\cite{FRT:lie}. We return to this in Section~5.

This describes the algebra. The novel braided-matrix multiplication property in
this R-matrix setting is that if $\vecu,\vecu'$ obey (\ref{B(R)})
then\cite{Ma:exa}
\eqn{mult}{ \vecu''=\vecu\vecu',\quad
R^{-1}\vecu_1'R\vecu_2=\vecu_2R^{-1}\vecu_1'R,\quad
{\rm i.e.}\quad  u_J'u_L= u_Ku_I' \Rmult ^I{}_J{}^K{}_L }
obeys (\ref{B(R)}) again. The novel braided-matrix addition property in this
setting is that also\cite{Mey:new}
\eqn{addn}{\vecu''=\vecu+\vecu',\quad  R^{-1}\vecu_1'R\vecu_2= \vecu_2
R_{21}\vecu_1'R,\quad
{\rm i.e.}\quad u_J'u_L= u_Ku_I' \Radd ^I{}_J{}^K{}_L }
obeys (\ref{B(R)}) as well.

This approach works in some generality: the multiplicative structure for any
bi-invertible solution $R$ of the QYBE and the additive one for any Hecke
solution. In the example of $q$-Minkowski space, we see that its basic `ring'
structure of addition and multiplication is described by three $16\times 16$
R-matrices: $\Rrel $ for the relations of $B(R)$, $\Rmult $ for the
multiplicative braid statistics and $\Radd $ for the additive braid statistics.

Finally, as well as introducing these ideas, examples, and general R-matrix
constructions for them\cite{Ma:exa}\cite{Mey:new}, we have a sound mathematical
foundation for them under the concept of a {\em braided-Hopf algebra} or
`braided group'. The formalisation of the braided-multiplication property is a
braided-coproduct
\[ \comult\vecu=\vecu\tens\vecu,\quad \Psi_\bullet(R^{-1}\vecu_1\tens
R\vecu_2)=\vecu_2R^{-1}\tens \vecu_1' R\]
and the formalisation of the braided-addition is a `braided-coaddition'
\[  \coadd\vecu=\vecu\tens 1+1\tens\vecu,\quad \Psi_+(R^{-1}\vecu_1\tens
R\vecu_2)=\vecu_2R_{21}\tens \vecu_1' R.\]
The $\tens$ in the two cases is no ordinary tensor product algebra but a
non-commuting one with relations between the two copies (as above)
expressed mathematically by the braid-transposition operator $\Psi$. So the
braided matrices (\ref{B(R)}) form a braided group in two ways,
one\cite{Ma:exa} with $\comult$ and braiding $\Psi_\bullet$ (and no antipode)
and the other\cite{Mey:new} with $\coadd$ and the braiding $\Psi_+$ and
braided-antipode $S\vecu=-\vecu$. This mathematical formulation of braided
groups is rather powerful as it allows all constructions to be done in a
systematic way using braid and tangle diagrams. See \cite{Ma:introp} for an
introduction to this novel technique.

This comultiplication and coaddition provided by (\ref{mult})--(\ref{addn}) are
the new feature of the braided-matrix approach to $q$-Minkowski space. Thus the
element $\und{\rm det}$ was introduced in \cite{Ma:exa} as the {\em braided
determinant} and is characterised by being multiplicative. Likewise, $t$ is the
{\em braided trace} and is characterised by being additive:
\[ \comult \und{\rm det}=\und{\rm det}\tens \und{\rm det},\quad \coadd t=t\tens
1+1\tens t.\]
In the interpretation of this algebra of $2\times 2$ braided matrices as
$q$-Minkowski space $\mink $, we clearly should take $\und{\rm det}$ as the
metric or norm. In a basis
\[ t=q^{-1}a+qd,\quad x={b+c\over 2},\quad y={b-c\over 2i},\quad z=d-a\]
this is
\[ \und{\rm det}={q^2\over(q^2+1)^2}t^2-q^2x^2-q^2 y^2-
{(q^4+1)q^2\over 2(q^2+1)^2}z^2+\left({q^2-1\over q^2+1}\right)^2{q\over 2}
tz.\]

Also from the theory of braided-geometry comes a natural $*$-structure
characterised by
\[ (*\tens *)\circ\comult=\tau\circ\comult\circ *,\quad (*\tens
*)\circ\coadd=\coadd\circ*\]
where $\tau$ is permutation. It is given by $\vecu$ Hermitean or equivalently,
$t,x,y,z$ are real in the sense $t^*=t$ etc. Again, this is a general feature
that works for any R-matrix of real-type in the sense $\overline
R=R_{21}^{t\tens t}$, which holds in the present case for real $q$. The
analysis is in \cite{Ma:mec}.

Next, the R-matrix ${\bf R}$ determines equally well a quantum {\em vector}
algebra with generators $v^I$ and a braided-addition law\cite{Ma:poi}. It has
relations
and additive braid statistics
\ceqn{coB(R)}{  v^Iv^K=\Rrel ^I{}_J{}^K{}_L v^Lv^J,\quad v'{}^Iv^K=\Radd
^I{}_J{}^K{}_L v^Lv'{}^J.}
This leads to a {\em quantum metric} $g^{IJ}$ with inverse $g_{IJ}$ and
providing an isomorphism
\cite{Mey:new}
\eqn{isom}{ v^I=g^{IJ}u_J,\quad u_I=g_{IJ}v^J}
between the vectors and covectors. It also obeys
\eqn{gsymm}{\Rrel^I{}_J{}^K{}_L g^{JL}=g^{KI},\quad
g_{IK}\Rrel^I{}_{J}{}^K{}_L=g_{LJ},\quad \und{\rm det}=u_Iu_Jg^{IJ}.}

The above structure then leads to a natural $q$-Lorentz group. We take the
quantum matrix FRT bialgebra  $A(\Radd)$ and, denoting the quantum matrix
generator by $\Lambda$, we define the quantum group $O_q(1,3)$ as this
bialgebra modulo the relation
\eqn{Lor}{ \Lambda^I{}_J\Lambda^K{}_L\, g^{JL}=g^{IK}.}
This $q$-Lorentz group coacts covariantly on the algebra (\ref{B(R)}) and its
vector version
\eqn{loract}{ \vecu\mapsto \vecu\Lambda,\quad \vecv\mapsto
\Lambda^{-1}\vecv,\quad {\rm i.e.}\quad u_J\mapsto u_I\tens \Lambda^I{}_J,\quad
v^I\mapsto v^J\tens S\Lambda^I{}_J. }
This $q$-Lorentz group respects the algebra $B(R)$ but not yet is additive
structure $\coadd$. According to the general scheme in \cite{Ma:poi}, {\em in
order for the braided addition of 4-vectors to be covariant} one must extend
the $q$-Lorentz group by a dilatation element $\dila $ say, with
\eqn{dila}{ [\dila ,\Lambda^I{}_J]=0,\quad \Delta \dila =\dila \tens \dila }
which is to be included in the coaction. This is the origin of the dilatation
element in our approach\cite{Ma:poi} (where it was denoted by $g$). It is
needed whenever $\lambda\ne 1$ where $\lambda\Radd$ is the quantum group
normalisation as explained in \cite{Ma:poi}.

Because the additive structure is fully covariant under this extended
$\widetilde{O_q(1,3)}$, we can make at once a semidirect product Hopf algebra
$\widetilde{O_q(1,3)}\rcross \mink $ which is the $q$-Poincar\'e group in our
approach\cite[Theorem 6]{Ma:poi}. In the present case it comes out as generated
by covectors $\vecp$ say and $q$-Lorentz transformations, with quantum group
structure
\ceqn{poin}{ p_I\, \dila =\lambda^{-1}\dila  p_I,\quad
p_I\Lambda^J{}_K=\lambda\Lambda^J{}_Np_M\Radd^M{}_I{}^N{}_K\\
\Delta p_J=p_J\tens \Lambda^J{}_I\, \dila +1\tens p_I,\qquad \lambda=q^{-1}.}
We explained in \cite{Ma:poi} how this general form arises as the bosonisation
of any linear braided momentum group. Unlike previous attempts\cite{SWW:inh},
the bosonisation construction is very general and it includes the case we need
now. Finally, this $q$-Poincar\'e group coacts on $q$-Minkowski space by
\eqn{poinact}{\vecu\mapsto \vecu\Lambda \dila +\vecp,\quad {\rm i.e.}\quad
u_J\mapsto u_I\tens \Lambda^I{}_J \dila +1\tens p_J.}

Another general application of the braided theory is that infinitesimal
translation using again the addition law (\ref{addn}) defines at once the
operators of differentiation according to the general theory of \cite{Ma:fre}.
Thus
\eqn{diff}{ (\del^I
f)(\vecu')=\left(u_I^{-1}(f(\vecu+\vecu')-f(\vecu'))\right)_{\vecu=0}}
by which we mean to order the terms in $f(\vecu+\vecu')$ using the additive
braid statistics (\ref{addn}) to put all the $\vecu$ to the left, and then pick
out the coefficient of $u_I$. It was shown in \cite{Ma:fre} that these $\del^I$
always realise the vector algebra $v^I$ above.  There are also
right-differentials defined in a similar way by translation from the right.

\section{Detailed comparison with the approach of Carow-Watamura et al}

The main idea behind the approach of \cite{CWSSW:ten}--\cite{OSWZ:def} is that
of spinors. They considered two non-commuting copies of the 2-dimensional
quantum plane, say $x_i$ and $y_j$ and considered the properties of the
null-four vectors $X_I=x_{i_0}y_{i_1}$ say, where $I$ is a multi-index (they
used upper indices but to match with the above conventions we write lower
ones). These authors obtained in this way an approach based on two $16\times
16$ R-matrices ${}_I{\bf R}$ and  ${}_{II}{\bf R}$. The second was the Lorentz
quantum group R-matrix introduced in \cite{CWSSW:lor} as a product of 4
R-matrices
\[  {}_{II}{\bf R}^I{}_J{}^K{}_L=R^{i_0}{}_a{}^{k_1}{}_b
R^a{}_{j_0}{}^{k_0}{}_c R^{i_1}{}_d{}^b{}_{l_1}
R^{-1}{}^c{}_{l_0}{}^d{}_{j_1}\]
and the first was found by decomposing it into projectors and taking a
different combination of them.

We give here two ways to embed this pioneering work into the braided-matrix
picture above. The first is to give the required change of basis explicitly,
while the second is more abstract but works in general. The explicit change of
basis is provided by the spinor metric $\eps_{ij}$ which was one of the key
ingredients in \cite{CWSSW:lor} and which is given by
\[ \eps_{ij}=\pmatrix{0&{1\over\sqrt{q}}\cr -{\sqrt{q}} &0} \]
in the relevant conventions. Its inverse is $\eps^{ij}$. We  then define
\ceqn{isomWess}{ X_I=\eps_{a i_0}u^a{}_{i_1},\quad T^I{}_J=\eps_{a
j_0}\Lambda^{a i_1}{}_{b j_1}\eps^{bj_0}\\
{\rm i.e.}\qquad X_J=u_IE^I{}_J,\quad {\bf T}=E^{-1}\Lambda E,\quad
E^I{}_J=\eps_{i_0 j_0}  \delta^{i_1}{}_{j_1}}
and find by an explicit calculation that
\[ {}_I{\bf R}=(E^{-1}\tens E^{-1})\Rrel (E\tens E),\quad {}_{II}{\bf
R}=(E^{-1}\tens E^{-1})\Radd (E\tens E).\]
In fact, we recover here ${}_I{\bf R}$ in a 4 R-matrix form
\[ {}_{I}{\bf R}^I{}_J{}^K{}_L=R^{-1}{}^{k_1}{}_a{}^{i_0}{}_b
R^{-1}{}^{k_0}{}_c{}^b{}_{j_0} R^{i_1}{}_d{}^a{}_{l_1}
R^d{}_{j_1}{}^c{}_{l_0}\]
which corresponds to the $16\times 16$ matrix defined in \cite{OSWZ:def} in
terms of projectors.

We now give the relationship in a more conceptual way which, if desired,
extends to the  same level of generality as the braided approach above. We make
the comparison by embedding the various $q$-Lorentz groups into the $*$-quantum
group $SU_q(2)\dcross SU_q(2)$ and comparing their images there. In general the
{\em spinorial $q$-Lorentz group} is $A\bowtie A$ as generated by two quantum
matrices $\vecs,\vect$ say with their usual FRT relations of $A(R)$ and
cross-relations and $*$-structure
\[ R\vect_1\vecs_2=\vecs_2\vect_1 R,\quad t^i{}_j{}^*=Ss^j{}_i,\quad
s^i{}_j{}^*=St^j{}_i\]
The factorisation construction $\bowtie$ needed here is from \cite{Ma:mor}
where this quantum group was studied, while the $*$-structure is as used
in\cite{CWSSW:lor}. One can also work at the bialgebra level by working with
\eqn{tdagger}{ \vect^\dagger\equiv S\vecs}
as the abstract generator in place of $\vecs$. So $A\bowtie A$ is generated by
$\vect^\dagger,\vect$ with appropriate relations and $*$-structure. We gave the
$*$-structure more abstractly in \cite[Sec. 4]{Ma:poi} and also that there is a
dual-quasitriangular structure or `universal R-matrix functional'
\eqn{uniRmult}{ \CR_M((a\tens b)\tens (c\tens d))=\CR^{-1}(d\o\tens
a\o)\CR^{-1}(c\o\tens a\t)\CR(b\o\tens d\t)\CR(b\t\tens c\t)}
for all $a,b,c,d\in A$ (this is actually the canonical inverse-transpose of the
one in \cite{Ma:poi}). We used the usual $a\o\tens a\t$ notation for the
coproduct. We also showed that this quantum group was the dual of the `twisted
square' in \cite{ResSem:mat} and hence in some nice cases isomorphic to the
dual of the quantum double\cite{Dri} as in the approach of \cite{PodWor:def}.
To this we add the observation that this bialgebra has a {\em second}
dual-quasitriangular structure\cite{Mey:new}
\eqn{uniRadd}{\CR_L((a\tens b)\tens (c\tens d))=\CR^{-1}(d\o\tens
a\o)\CR(a\t\tens c\o)\CR(b\t\tens c\t)\CR(b\o\tens d\t).}
In these formulas, $\CR$ is the dual quasitriangular structure of $A$ and is
real in the sense of \cite[Prop. 13]{Ma:poi} as is the case for $SU_q(2)$ at
real $q$. Then (\ref{uniRadd}) is also real while (\ref{uniRmult}) is
anti-real.

The vectorial $q$-Lorentz group of Section~2 then maps into this spinorial one
by
\eqn{spinrep}{ \Lambda^I{}_J=t^\dagger{}^{j_0}{}_{i_0} t^{i_1}{}_{j_1}}
giving the transformation law
\eqn{spinrepcoact}{u^i{}_j\mapsto u^a{}_b t^\dagger{}^i{}_a t^b{}_j,\qquad {\rm
i.e.},\quad\vecu \mapsto
\vect^\dagger\vecu\vect}
while the original embedding of \cite{CWSSW:lor} is
\[ T^I{}_J=s^{i_0}{}_{j_0}t^{i_1}{}_{j_1}, \qquad X_{ij}\mapsto X_{ab} s^a{}_i
t^b{}_j.\]
Moreover, one has consistency in that the above dual-quasitriangular structures
recover the specific R-matrices above. Thus
\[ \Rrel=\CR_M(\Lambda_1\tens\Lambda_2),\quad
\Radd=\CR_L(\Lambda_1\tens\Lambda_2)\]
\[ {}_I{\bf R}=\CR_M(T_1\tens T_2),\quad {}_{II}{\bf R}=\CR_L(T_1\tens T_2)\]
This is a general construction, but comparing the specific realisations of $T$
and $\Lambda$ in this quantum group $A\dcross A$ gives the matrix $E$ or the
spinor metric $\eps$ for the standard case.

In summary, one could say that the braided-matrix picture of $q$-Minkowski
space and the original spinor picture differ by a change of basis. The latter
spinorial approach adds the spinor metric (we do not need it in the braided
matrix approach) while the braided matrix approach adds the concepts of braided
comultiplication (which in turn determines the metric) and braided coaddition.
They come out as $\coadd X=X\tens 1+1\tens X$ (or $X+X'$ where the two copies
have the additive braid statistics ${}_{II}{\bf R}$), and
\eqn{mult-X}{ \comult X_{ij}=X_{ia}\tens \eps^{ab} X_{bj}}
with respect to multiplicative braid statistics
\[ {}_{III}{\bf R}=(E^{-1}\tens E^{-1})\Rmult(E\tens E),\quad {}_{III}{\bf
R}^I{}_J{}^K{}_L=R^{-1}{}^{k_1}{}_a{}^{i_0}{}_b R^{i_1}{}_c{}^a{}_{l1}
R^{-1}{}^{k_0}{}_d{}^b{}_{j_0}R^c{}_{j_1}{}^d{}_{l0}.\]
Moreover, the braided approach avoids the use of projectors and thereby stays
in a general R-matrix form.

\section{Spinorial $q$-Poincar\'e group}

We now continue with further results in the braided-matrix approach. By using
the change of basis provided in Section~3, our results can be viewed equally
well as new results in the approach of \cite{CWSSW:ten}--\cite{OSWZ:def}. In
fact, the basis $\{u^i{}_j\}$ has a number of advantages over the $\{X_{ij}\}$
basis whenever braided-matrix multiplication (\ref{mult}) is needed implicitly
or explicitly. A general rule is that for vector multi-index descriptions
either basis is fine but the $\vecu$-basis has an advantage if we really want
to work with the original spinorial $4\times 4$ R-matrix.

Here we apply the techniques of braided geometry\cite{Ma:introp} to obtain the
$q$-Poincar\'e group (\ref{poin}) in the spinorial representation
(\ref{spinrep}) for the $q$-Lorentz group part. This gives a picture of the
$q$-Poincar\'e group as generated by two copies $\vecs,\vect$ of $SU_q(2)$ and
the momentum braided group of translations as a copy of $q$-Minkowski space
with its additive structure (\ref{addn}). At issue is the commutation relations
between $p_I$ and the $\vecs,\vect$. In the braided-matrix approach this
follows as a new application of the general {\em bosonisation theorem} in
\cite{Ma:poi}. The idea comes from the Jordan-Wigner transform whereby a
super-matrix is realised as an ordinary matrix by `bosonising'.

Firstly, we extend the spinorial $q$-Lorentz group by a dilatation element
$\dila $ as in (\ref{dila}), commuting with $\vecs,\vect$. We also extend the
dual-quasitriangular structure $\CR_L$ in (\ref{uniRadd}) by $\CR_L(\dila
^a\tens \dila ^b)=\lambda^{-ab}$. This describes the extended quantum group
$\widetilde{SU_q(2)\bowtie SU_q(2)}$ which is a double-cover of the vector
Lorentz group $\widetilde{O_q(1,3)}$ in Section~2. It coacts as in
(\ref{spinrepcoact}) with an extra  $\dila$ factor.

Then one can show that $q$-Minkowsi space is a braided-Hopf algebra fully
covariant under this extended spinorial $q$-Lorentz group. Its braiding is that
induced by $\CR_L$. Hence by the bosonisation theorem as applied in
\cite[Theorem~6]{Ma:poi} we obtain an ordinary Hopf algebra $q$-Poincar\'e
group as a semidirect product $\left(\widetilde{SU_q(2)\bowtie
SU_q(2)}\right)\rcross\mink $. Following the same steps to compute this as in
\cite{Ma:poi} we first define the right action of the extended $q$-Lorentz
group on $\mink $ by evaluating $\CR_L$ against its coaction. This gives
\cmath{ p^i{}_j\ra t^k{}_l=p^a{}_b\CR_L\left((Ss^i{}_a) t^b{}_j \dila \tens
t^k{}_l\right)=p^a{}_b\CR^{-1}(t^k{}_c\tens St^i{}_a)\CR(t^b{}_j\tens
t^c{}_l)=\lambda p^a{}_bR^k{}_c{}^i{}_aR^b{}_j{}^c{}_l\\
p^i{}_j\ra s^k{}_l=p^a{}_b\CR_L\left((Ss^i{}_a) t^b{}_j \dila \tens
s^k{}_l\right)=p^a{}_b\CR(St^i{}_a\tens t^k{}_c)\CR(t^b{}_j\tens t^c{}_l)=
p^a{}_bR^{-1}{}^i{}_a{}^k{}_c R^b{}_j{}^c{}_l\\
{\rm i.e.}\quad \vecp_1\ra \vect_2=\lambda R_{21}\vecp_1 R,\quad  \vecp_1\ra
\vecs_2= R^{-1}\vecp_1 R.}
This right action gives the semidirect product algebra structure of the
$q$-Poincar\'e group, while the coaction itself gives the semidirect coproduct.
The result is that our spinorial $q$-Poincar\'e group has the cross-relations
and coproduct
\eqn{spinpoin1}{\vecp_1\vect_2=\lambda\vect_2R_{21}\vecp_1 R,\quad
\vecp_1\vecs_2=\vecs_2 R^{-1}\vecp_1 R,\quad \vecp \dila =\lambda^{-1}\dila
\vecp}
\eqn{spinpoin2}{\Delta \vecp=\vecp\tens\vect^\dagger(\ )\vect \dila
+1\tens\vecp,\quad {\rm i.e.}\quad \Delta p^i{}_j=p^a{}_b\tens
(Ss^i{}_a)t^b{}_j \dila +1\tens p^i{}_j}
where $\vect^\dagger(\ )\vect$ has a space for the matrix indices of $\vecp$ to
be inserted. There is also a counit $\eps \vecp=0$ and antipode. As far as we
know, this R-matrix construction is new. It is the double-cover of the
vectorial $q$-Poincar\'e group in Section~2. It coacts on $q$-Minkowski space,
which becomes a comodule algebra as in (\ref{poinact}) but in the spinor
version with (\ref{spinrep}).

We have worked in an R-matrix formulation and quantum groups of function
algebra type, but it is not hard to give the enveloping algebra version too by
the same techniques. Thus the vector algebra (\ref{coB(R)}) is represented by
the $\del^I$ and is dually-paired as a braided-Hopf algebra with the
$\{p_I\}$\cite{Ma:poi}, the pairing
being in the categorical sense explained in \cite[Prop. 4.11]{Ma:introp}. The
spinorial Lorentz group has enveloping algebra $U_q(su_2)\tens_{\CR} U_q(su_2)$
along the lines of \cite{ResSem:mat} with $*$-structure and $\CR_L$ dual to
those above. Its extension  $U_q(su_2)\tens_{\CR} U_q(su_2)\tens U(u(1))$ acts
on the vector algebra and cross product by this (in other words the
bosonisation of the vector algebra) constructs the enveloping algebra of the
$q$-Poincar\'e group. By general theorems cf\cite[Lemma~4.4]{Ma:mec} it will be
a Hopf algebra and dual to the $q$-Poincar\'e group above. Details will be
presented elsewhere.

Another variant is to work with right-handed derivatives
$\overleftarrow{\del^I}$ rather than usual left ones. Related to this, it is
obvious that all formulae could be presented with upper and lower indices
swapped (there are some reasons for our conventions above which do not show up
in the classical case where algebras of functions are commutative.)

Finally, if one wants to stay entirely in the braided setting, there is a
braided $q$-Lorentz group $BO_q(1,3)$ based on braided matrices $B(\Radd)$
rather than quantum matrices $A(\Radd)$ as above. The braided-semidirect
product\cite[Eq. 72]{Ma:introp} of $q$-Minkowski space by this gives the
braided-Poincar\'e group. Its advantage over the one above is that the
translation part appears as a sub-braided group. This too will be computed
elsewhere.

\section{Concluding remarks}

We conclude with some comments about more novel aspects of our braided-matrix
approach. Firstly, it is very general and applies just as well to other
real-type Hecke R-matrices. For example, a non-standard variant of
(\ref{Rjones}) gives the braided version $BM_q(1|1)$ of super-$2\times 2$
matrices introduced in \cite{Ma:exa}. It provides a super-version of
$q$-Minkowski space with some interesting features such as vanishing
quantum-dimension.

Secondly, there is by now a systematic theory of braided-Lie algebras
introduced in \cite[Prop.~2.4]{Ma:skl}\cite{Ma:lie} and associated to any
bi-invertible R-matrix. The one corresponding to (\ref{Rjones}) is computed in
\cite[Ex.~5.5]{Ma:lie} as braided $gl_{2,q}$. It turns out that the canonical
braided-enveloping algebra for this class of examples
is isomorphic when $q\ne 1$ to the braided-matrix bialgebras $B(R)$ with their
multiplicative coproduct. A part of this (if one sets $\und{\rm det}=1$ etc) is
the known picture of (\ref{B(R)}) as a description of quantum enveloping
algebras\cite{ResSem:cen}\cite{FRT:lie} but in the braided theory we do not
have to make such restrictions. Thus for example, we have
\[ U(gl_{2,q})\isom\mink \]
as $*$-algebras. The non-commuting space-time co-ordinates $t,x,y,z$ can be
viewed equally well on the left hand side as generators of $gl_{2,q}$. The time
direction $t$ corresponds to the $u(1)$ direction. This remarkable possibility
of the unification of a $SU(2)\times U(1)$ symmetry in the form of $gl_{2,q}$
with the co-ordinates of space-time itself
is perhaps the main result of \cite{Ma:lie} for physics. Its meaning was
discussed further in \cite{Ma:mex}.

Thirdly, just as the braided matrices $B(R)$ have these two interpretations, so
the $q$-Lorentz group above has at least {\em three} interpretations. One is as
above, another is as the quantum algebra of observables of a $q$-deformed
particle moving on a hyperboloid in $q$-Minkowski space, which is the main
result of \cite{Ma:mec}. A third is as a `frame bundle' in quantum group gauge
theory\cite[Ex.~5.6]{BrzMa:gau}. These are all valid pictures of Drinfeld's
quantum double within quantum and braided geometry. In summary, one can say
that {\em $q$-deformation unifies concepts} because structures which are quite
different at $q=1$ can become isomorphic when $q\ne 1$. This provides therefore
a novel motivation for $q$-deforming physics.


\begin{thebibliography}{10}
\itemsep 0pt


\bibitem{CWSSW:ten}
U.~Carow-Watamura, M.~Schlieker, M.~Scholl, and S.~Watamura.
\newblock Tensor representation of the quantum group {$SL_q(2,\C)$} and quantum
  {M}inkowski space.
\newblock {\em Z. Phys. C}, 48:159, 1990.



\bibitem{CWSSW:lor}
U.~Carow-Watamura, M.~Schlieker, M.~Scholl, and S.~Watamura.
\newblock A quantum {L}orentz group.
\newblock {\em Int. J. Mod. Phys. A}, 6:3081--3108, 1991.




\bibitem{OSWZ:def}
O.~Ogievetsky, W.B. Schmidke, J.~Wess, and B.~Zumino.
\newblock {$q$}-Deformed {P}oincar{\'e} algebra.
\newblock {\em Comm. Math. Phys.}, 150:495--518, 1992.

\bibitem{Ma:introp}
S.~Majid.
\newblock Beyond supersymmetry and quantum symmetry (an introduction to braided
  groups and braided matrices).
\newblock In M-L. Ge and H.J. de~Vega, editors, {\em Quantum Groups, Integrable
  Statistical Models and Knot Theory}, pages 231--282. World Sci., 1993.

\bibitem{Ma:exa}
S.~Majid.
\newblock Examples of braided groups and braided matrices.
\newblock {\em J. Math. Phys.}, 32:3246--3253, 1991.

\bibitem{Ma:mec}
S.~Majid.
\newblock The quantum double as quantum mechanics, {S}eptember 1992.
\newblock To appear in {\em J. Geom. Phys.}


\bibitem{Mey:new}
U.~Meyer.
\newblock A new {$q$}-{L}orentz group and $q$-{M}inkowski space with both
  braided coaddition and {$q$}-spinor decomposition.
\newblock {\em Preprint}, DAMTP/93-45, 1993.

\bibitem{Ma:poi}
S.~Majid.
\newblock Braided momentum in the {$q$}-{P}oincar{\'e} group.
\newblock {\em J. Math. Phys.}, 34:2045--2058, 1993.

\bibitem{Ma:lin}
S.~Majid.
\newblock Quantum and braided linear algebra.
\newblock {\em J. Math. Phys.}, 34:1176--1196, 1993.

\bibitem{Ma:skl}
S.~Majid.
\newblock Braided matrix structure of the {S}klyanin algebra and of the quantum
  {L}orentz group.
\newblock {\em Comm. Math. Phys.}, 156:607--638, 1993.

\bibitem{ResSem:cen}
N.Yu. Reshetikhin and M.A. Semenov-Tian-Shansky.
\newblock Central extensions of quantum current groups.
\newblock {\em Lett. Math. Phys.}, 19:133--142, 1990.

\bibitem{FRT:lie}
L.D. Faddeev, N.Yu. Reshetikhin, and L.A. Takhtajan.
\newblock Quantization of {L}ie groups and {L}ie algebras.
\newblock {\em Leningrad Math J.}, 1:193--225, 1990.

\bibitem{SWW:inh}
M.~Schlieker, W.~Weich, and R.~Weixler.
\newblock Inhomogeneous quantum groups.
\newblock {\em Z. Phys. C}, 53:79--82, 1992.

\bibitem{Ma:fre}
S.~Majid.
\newblock Free braided differential calculus, braided binomial theorem and the
  braided exponential map.
\newblock {\em J. Math. Phys.}, 34:4843--4856, 1993.

\bibitem{Ma:mor}
S.~Majid.
\newblock More examples of bicrossproduct and double cross product {H}opf
  algebras.
\newblock {\em Isr. J. Math}, 72:133--148, 1990.

\bibitem{ResSem:mat}
N.Yu. Reshetikhin and M.A. Semenov-Tian-Shansky.
\newblock Quantum {$R$}-matrices and factorization problems.
\newblock {\em J. Geom. Phys.}, 5:533, 1988.

\bibitem{Dri}
V.G. Drinfeld.
\newblock Quantum groups.
\newblock In A.~Gleason, editor, {\em Proceedings of the {ICM}}, pages
  798--820, Rhode Island, 1987. AMS.

\bibitem{PodWor:def}
A.~Podles and S.L. Woronowicz.
\newblock Quantum deformation of {L}orentz group.
\newblock {\em Comm. Math. Phys}, 130:381--431, 1990.

\bibitem{Ma:lie}
S.~Majid.
\newblock Quantum and braided {L}ie algebras, {F}ebruary 1993.
\newblock To appear in {\em J. Geom. Phys.}


\bibitem{Ma:mex}
S.~Majid.
\newblock Lie algebras and braided geometry.
\newblock To appear in {\em Proc. XXII's DGM, Ixtapa, Mexico, September, 1993}


\bibitem{BrzMa:gau}
T.~Brzezi\'nski and S.~Majid.
\newblock Quantum group gauge theory on quantum space.
\newblock {\em Comm. Math. Phys.}, 157:591--638, 1993.

\end{thebibliography}

\end{document}